\documentclass[prl,aps,epsf,twocolumn,superscriptaddress,longbibliography]{revtex4-2}
\usepackage{graphicx}
\usepackage{overpic}
\usepackage{dcolumn}
\usepackage{bm}
\usepackage{color}
\usepackage{soul}
\usepackage{hyperref}
\usepackage{amssymb}
\usepackage{amsmath}
\usepackage{wasysym}
\usepackage{tabularx}

\usepackage{appendix}
\usepackage{hyperref}
\usepackage{float}
\usepackage[mathscr]{euscript}
\usepackage[scr=boondox]{mathalpha}

\newcommand{\SO}{\mathscr{S}_{\text{o}}}
\newcommand{\OO}{\text{o}}
\newcommand{\PO}{\vec{\mathscr{P}}_{\OO}}

\begin{document}
\preprint{APS/123-QED}

\title{Discrete Shift and Polarization from Response to Symmetry Defects in Interacting Topological Phases}

\author{Lu Zhang}
\thanks{These authors contributed equally}
\affiliation{%
 The Department of Physics, Hong Kong University of Science and Technology, Clear Water Bay, Hong Kong, China}
 
 \author{Min Long}
 \thanks{These authors contributed equally}
\affiliation{Department of Physics and  HK Institute of Quantum Science \& Technology, The University of Hong Kong, Pokfulam Road,  Hong Kong SAR, China}

 \author{Yuxuan Zhang}
 \affiliation{Department of Physics and Joint Quantum Institute, University of Maryland, College Park, Maryland 20742, USA}

 \author{Zi Yang Meng}
\affiliation{Department of Physics and  HK Institute of Quantum Science \& Technology, The University of Hong Kong, Pokfulam Road,  Hong Kong SAR, China}

\author{Xue-Yang Song}%
\affiliation{The Department of Physics, Hong Kong University of Science and Technology, Clear Water Bay, Hong Kong, China} 
\date{\today}

\begin{abstract}
    We extend the previous study of extracting crystalline symmetry-protected topological invariants to the correlated regime. We construct the interacting Hofstadter model defined on square lattice with the rotation and translation symmetry defects: disclination and dislocation.  The model realizes Chern insulator and the charge density wave state as one tunes interactions.  Employing the density matrix renormalization group (DMRG) method, we calculate the excess charge around the defects and find that the topological invariants remain quantized in both phases, with the topological quantity extracted to great precision. 
    This study paves the way for utilizing matrix product state, and potentially other quantum many-body computation methods, to efficiently study crystalline symmetry defects on 2D interacting lattice systems.
\end{abstract}

\maketitle

\noindent{\textcolor{blue}{\it Introduction}---}The classification of topological phases of matter can be characterized by quantized response invariants. 
The interplay between symmetry and topology leads to a broad landscape of symmetry-protected topological (SPT) phases, and many-body interactions can further refine or alter the classification.  These issues have been extensively discussed for internal and crystalline symmetries, at the single-particle level, as well as in interacting systems ~\cite{kitaev2009periodic,ryu2010topological,hohenadlerQuantum2012,chiu2016classification,chen2013symmetry,kapustin2014symmetry,kapustin2015fermionic,gu2014symmetry,else2014classifying,wang2014interacting, senthil2015symmetry,heBonafide2016,heTopologicalI2016,heTopologicalII2016,song2017interaction,thorngren2018gauging,barkeshli2022classification,aasen2021characterization,zhang2022construction}. 
One paradigmatic example is the integer quantum Hall effect that preserves the $\text{U}(1)$ symmetry~\cite{ando1975theory}. 
Distinct topological phases are characterized by different Chern number $C$, which gives rise to the quantized charge Hall response.  
From another point of view, this topological invariant indicates that there are $C$ charges nucleated by one $\text{U}(1)$ flux, according to the Streda formula $\Delta n = C \Delta \phi$. 
This response serves as the topological invariant, which remains unchanged as one moves inside the same SPT phase.

Beyond $\text{U}(1)$ charge symmetry, the topological response to the flux of spatial symmetry, encoded in geometric curvature and torsion~\cite{avron1995viscosity,read2011hall,haldane2009hall,haldane2011geometrical,abanov2014electromagnetic,gromov2015framing} has also been proposed.
On the lattice, continuous spatial symmetries such as the rotation and translation symmetry reduce to crystalline symmetries, e.g., discrete rotations $C_M$ and lattice translations $T_{x(y)}$. The crystalline symmetry enriches the classification of the $\text{U}(1)$ symmetry protected topological order~\cite{thorngren2018gauging}.
In close analogy with $\text{U}(1)$ flux insertion in the $\text{U}(1)$ SPT case, crystalline symmetries admit topological defects: disclination and dislocation, characterized by the disclination angle  $\Omega$ and the Burger's vector $\vec b$~\cite{liu2019shift,benalcazar2014classification,li2020fractional,you2020higher,PhysRevB.106.L241113,cspt_nature} respectively.
The corresponding $\text{U}(1)$ charge responses to crystalline symmetry defects give rise to new topological invariants discrete shift $\SO$ and polarization $\PO$, where $o$ is chosen as the high symmetry points of the unit cell, that diagnose distinct crystalline SPT phases.


Previous studies on the response to the symmetry defects focused on single-particle Hamiltonian~\cite{zhang2022fractional,zhang2023quantized} or fine-tuned wavefunctions~\cite{zhang2024fractionally,kobayashi2024crystalline}.
It nevertheless remains elusive to obtain the response in \emph{interacting} systems. 
This work reports a numerical study to diagnose the crystalline symmetry protected order in correlated systems, in the framework of matrix product state(MPS).

We study a minimal lattice model that realizes nontrivial topology: the Hofstadter model on the square lattice with nearest-neighbor interactions. We construct many-body Hamiltonian on lattice that host crystalline symmetry defects—disclinations and dislocations—thereby enabling directly probing of their associated topological responses. Using large-scale density matrix renormalization group(DMRG) calculations~\cite{White1992Density,Schollwck2011DMRG}, we obtain real-space charge density distributions and extract the defect-bound charge. The measured responses are quantized and match the theoretical predictions for the discrete shift and polarization, providing nonperturbative evidence for the robustness of these crystalline response invariants. Remarkably,
we tune the interaction strength that drives the system into the charge density wave(CDW) phase, and find that the CDW phase also acquires the quantized response of the discrete shift, even without band topology. These results demonstrate that the topological invariants are still well-defined in the strongly-correlated regime beyond the single particle band analysis.  
To our knowledge, this work presents the first DMRG study of disclination- and dislocation-induced responses in interacting systems, and highlights MPS and potentially other quantum many-body numerical methods as controlled tools for investigating crystalline defects and their topological responses.
\begin{figure}[h]
    \centering
    \includegraphics[width=1.0\linewidth]{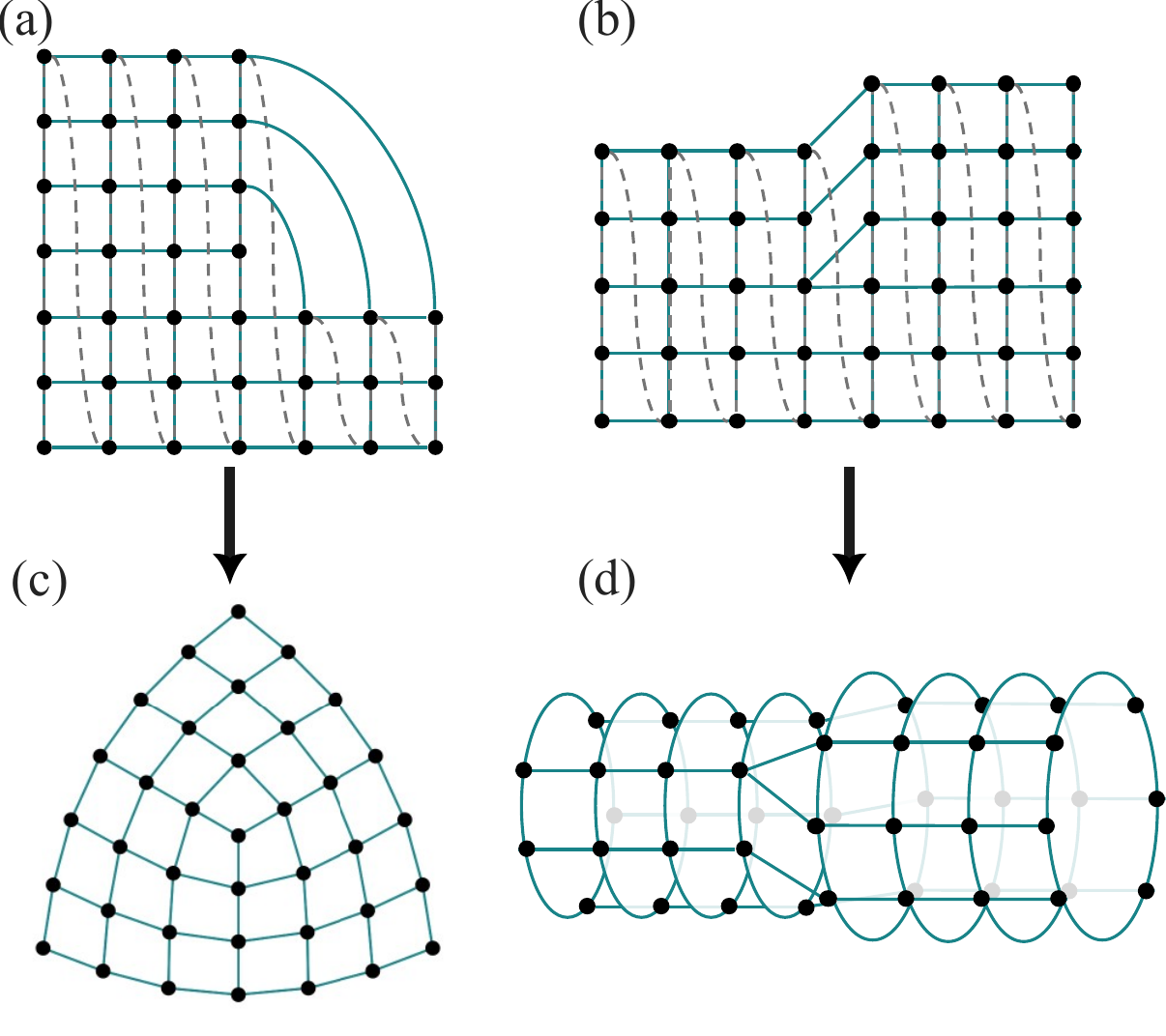}
    \caption{The disclination (a,c) with $\Omega = \frac{\pi}{2}$ and  dislocation (b,d) with $\vec b = (0,-1)$  constructed using a cut-and-glue procedure (see SI~\cite{suppl} for details). The green solid lines denote the hopping of electrons while the gray dashed lines illustrate the "snake" of the one-dimensional matrix product state. In the case of dislocation, we apply periodic boundary condition in $y$ to eliminate boundary effects.}
    \label{fig:lattice}
\end{figure}

\noindent{\textcolor{blue}{\it Model Hamiltonian and topological field theory}---}We consider spinless fermions on a square lattice in a uniform magnetic field, described by the interacting Hofstadter Hamiltonian
\begin{equation}
    H = -t\sum_{\langle ij\rangle }e^{iA_{ij}}c_i^\dagger c_j + V\sum_{\langle ij\rangle} n_i n_j
    \label{eq:Hamiltonian}
\end{equation}
where $A_{ij}$ is a static background U$(1)$ gauge field that yields a flux $\phi$ per elementary plaquette. $c_i^\dagger (c_i)$ creates (annihilates) a spinless fermion at site $i$ and $n_i = c_i^\dagger c_i$ is the local density operator. $\langle ij\rangle$ runs over nearest neighbors.
In addition to the $\text{U}(1)$ conversation, the system stays invariant under $C_4$ rotation and translations along $x$ and $y$ directions, $T_x,T_y$.  With magnetic flux, the crystalline symmetry $\mathbb{Z}^2 \rtimes \mathbb{Z}_4$ becomes $\text{U}(1)\times_{\phi}\left(\mathbb{Z}^2 \rtimes \mathbb{Z}_4\right)$, i.e. centrally extended by $\text{U}(1)$\footnote{It means that the crystalline symmetry is realized projectively under the background magnetic field. For example, the translation symmetry generators $T_x$ and $T_y$ satisfies the following relation: $T_xT_yT_x^{-1}T_y^{-1}=e^{i\phi}$}.
In the following, we set $t = 1$ and $\phi = \frac{13}{12}\pi$ throughout unless specified otherwise and tune the interaction strength $V$ to investigate topological responses of the system. The filling is chosen to be commensurate with the magnetic flux and the system is in the gapped phase both in Chern insulator and CDW)(see SI~\cite{suppl}).  Without interaction, the electrons fills a complete Chern band with Chern number $C = -1$, and we denote it integer quantum Hall (IQH) state.

To describe the responses of the symmetry defects we introduce the symmetry gauge fields~\cite{song2021electric, manjunath2021crystalline} $A$, $A_\omega$ and $A_T$, corresponding to the $\text{U}(1)$ charge symmetry, rotation symmetry and translation symmetry respectively. The U($1$) charge response to the symmetry defect can be obtained from topological field theory:
\begin{equation}
    \mathcal{L}_{\text{CS}}=\frac{C}{4 \pi} A \wedge d A+\frac{\SO}{2 \pi} A \wedge d A_\omega+\frac{\vec{\mathscr{P}}_{\text{o}}}{2 \pi} A \wedge dA_T +\cdots,
    \label{eq:response}
\end{equation}
where $C$, $\SO$ and $\PO$ are the Chern number, discrete shift and polarization vector specifying the charge bound to the symmetry defects of $\text{U}(1)$ charge we choose the vertex site as $o$ in our simulation~\footnote{The discrete shift and polarization vector in lattice depends on the choice of the origin $o$, which is not the focus of the current paper. The details can be found in paper~\cite{zhang2023quantized}}), rotation and translation symmetries, respectively. ``Charge bound with symmetry flux'' picture can be obtained by examining mutual Chern-Simons term. For instance, considering the term $\frac{\SO}{2 \pi} A \wedge d A_\omega$, there should be $\SO$ charge attached to the $2\pi$ disclination. Mathematically, this can be derived by coupling system with the matter field($j_\mu A^\mu$)  and perform the variation of the Chern-Simons term over the $\text{U}(1)$ gauge field. We have omitted other crystalline terms in Lagrangian (Eq.~\eqref{eq:response}) which are not the concern of this study.
It can be shown theoretically that the discrete shift satisfies the condition: 
\begin{equation}
\SO = \frac{C}{2} \mod 1,
\label{eq:S_C}
\end{equation}
which means that the $\SO$ is (half)integer for (odd)even Chern number $C$.
The simple arguments of the quantization of discrete shift is provided in the supplemental information (SI)~\cite{suppl}.
For the system with the $C_4$ rotation, one can also show that the polarization vector $\PO$ can only take $(\frac{1}{2},\frac{1}{2})$ or $(0,0)$~\cite{zhang2023quantized}. In the following we use $\mathscr{P}_{o,y}$ to denote $y$ component of polarization vector.

\noindent{\textcolor{blue}{\it DMRG calculation}---}The field theory in Eq.~\eqref{eq:response} combined with the quantization of topological invariants predict the quantized charge bound to the symmetry defects. Upon turning on the interaction, the ground state can be obtained by DMRG simulation. In our simulation of disclination, we consider a square lattice with $13\times 13 $ sites and perform the cut-and-glue procedure shown in Fig.~\ref{fig:lattice} (a) and (c). While for dislocation, as shown in Fig.~\ref{fig:lattice} (b) and (d), the cylinder geometry with width $8$ is chosen. To avoid local minimum, we choose a multiple (around 20) initial ansatz and determine the ground state with the lowest energy. Details about the cut-and-glue procedure and DMRG simulation are presented in SI~\cite{suppl}. $\text{U}(1)$ symmetry is utilized in DMRG simulation.

To compute $\SO$ and $\PO$, we construct the lattice with corresponding symmetry defects(disclinations and dislocations). According to the response theory Eq.~\eqref{eq:response}, the Chern number is understood within the ``charge bound to magnetic flux'' picture, as described by the Streda formula. Similarly, the discrete shift $\SO$ and polarization vector $\PO$ can be interpreted in terms of ``charge bound to defects'', where they are derived from the excess charge $\delta Q$ bounded with disclinations and dislocations, as we now turn to.


\begin{figure}
    \centering
    \includegraphics[width=1.0\linewidth]{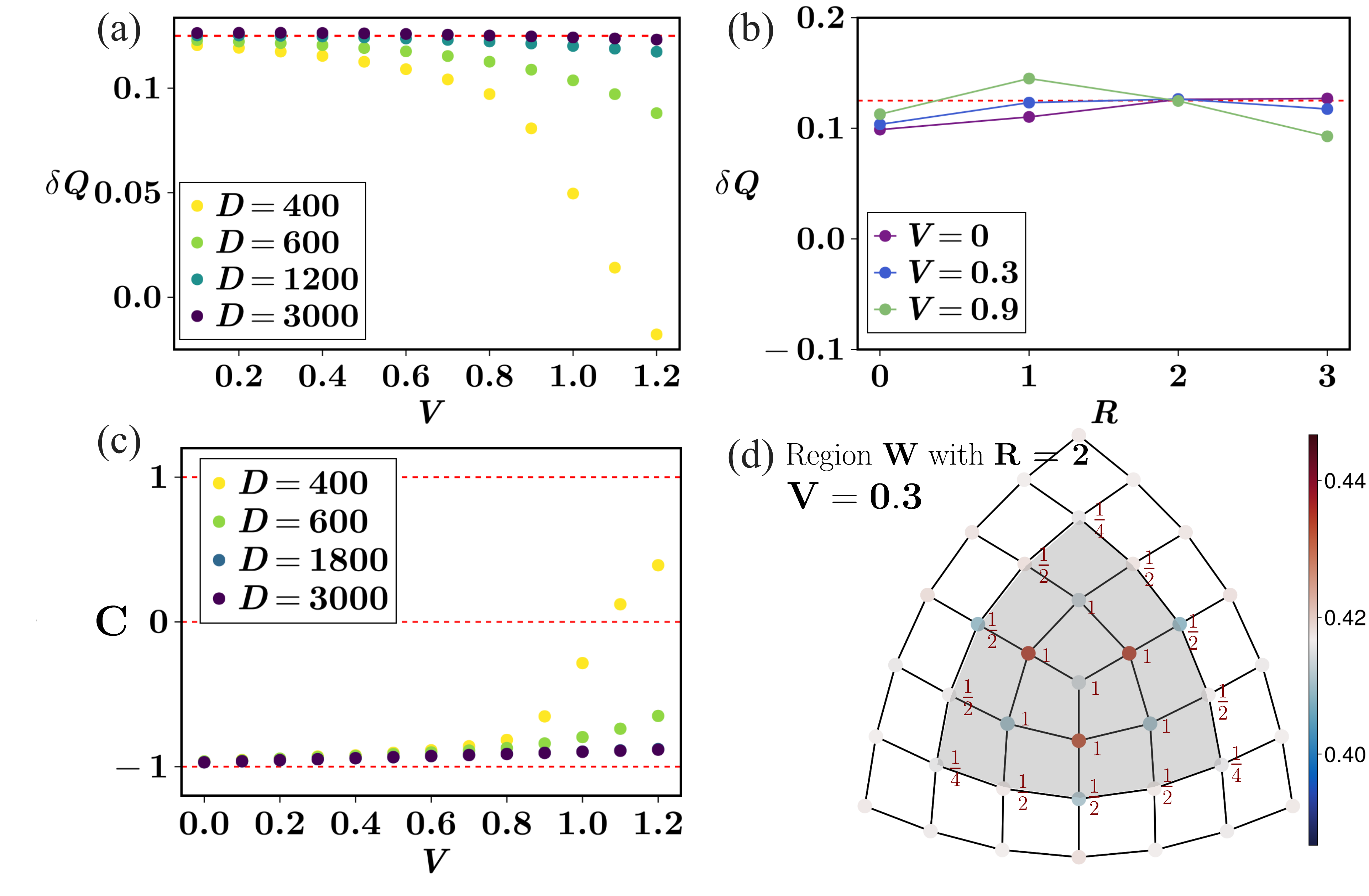}
    \caption{ (a) The excess charge measured around the disclination of different bond dimensions and interaction strengths. (b) The excess charge measured in regions with different sizes and different interaction strengths. The red dashed lines in (a) and (b) highlight the $1/8$ quantization of $\delta Q$. (c) The charge pumping by threading an infinitesimal flux at the disclination. The $C = \frac{dn}{d\phi}$ is calculated by measuring the change of the total charge inside the selected area, which gives the Chern number of the system according to the Streda formula.  The red dotted lines highlight the integer quantization (d) The density plot at $V = 0.3$. We only show the lattice away from the boundary. The shaded region $W$ with $R=2$ is where we choose to compute the excess charge $\delta Q$. The color on lattice sites labels the charge density according to the color bar.}
    \label{fig:disclination}
\end{figure}

\begin{figure}
    \centering
    \includegraphics[width=1.0\linewidth]{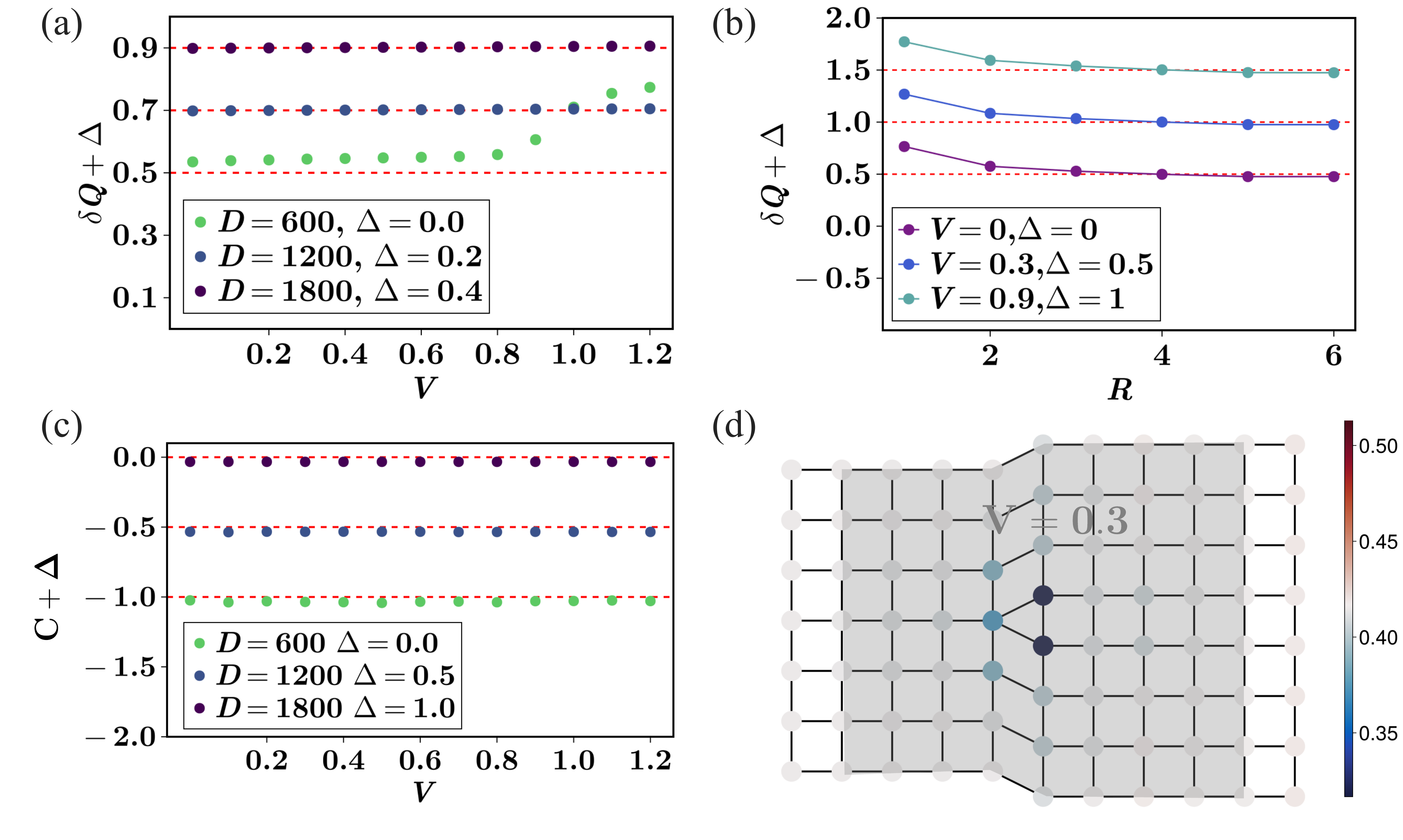}
        \caption{ With periodic boundary condition in $y$, the dislocation is put in a cylinder. We shift the data by $\Delta$ to show its convergence in (a) (b), and (c). (a) The excess charge measured around the dislocation of different bond dimensions for different interaction strengths. (b) The size scaling of excess charge for three different interaction strengths $V= 0,0.3,0.9$. (c) The charge pumping by threading an infinitesimal flux into the cylinder. The Chern number $C = \frac{dn}{d\phi}$ is calculated by measuring the change of the total charge of the left boundary.  (d) The density plot near the dislocation with $V = 0.3$. We choose the shaded region to compute the excess charge $\delta Q$. The color on lattice sites labels the charge density.}
    \label{fig:dislocation}
\end{figure}

\begin{table*}
    \centering
    \renewcommand\arraystretch{1.5}
    \tabcolsep = 0.24cm
    \begin{tabular}{|c|ccccccccccccc|}
    \hline 
               $V$   & 0.0    & 0.1   & 0.2   & 0.3   & 0.4   & 0.5   & 0.6   & 0.7   & 0.8   & 0.9   & 1.0   & 1.1   & 1.2 \\ \hline
      $\SO$  & 0.505 & 0.505 & 0.506 & 0.506 & 0.506 & 0.505 & 0.504 & 0.502 & 0.500 & 0.499 & 0.497 & 0.495 & 0.493 \\ 
      $\mathscr{P}_{o,y}$  & 0.498 & 0.499 & 0.500 & 0.501 & 0.501 & 0.502 & 0.502 & 0.503 & 0.504 &  0.504 &  0.505 &  0.505 &  0.506\\
      \hline
    \end{tabular}
    \caption{The discrete shift $\SO$ and polarization $\PO$ from DMRG calculation of bond dimension $m=3000$ and $m = 1800$. Both $\SO$ and $\mathscr{P}_{o,y}$ are found to be close to the expected value of $\frac{1}{2}$.}
    \label{tab:SO and PO}
\end{table*}

\noindent{\textcolor{blue}{\it Excess charge of disclination and dislocation}---}We first discuss disclination then move to dislocation. The charge responses to topological defects are measured in a finite region that encloses the defects. For disclination, the region $W$ is visualized in Fig.~\ref{fig:disclination}(d) with its size labeled by $R$. $R$ is chosen such that its boundary is far from the boundary and the symmetry defects. The center of the region $W$ is located at the vertex of lattice, which is the choice of our origin $o$. The total charge inside the region $R$ is $Q_W = \sum_{i\in W} \text{w}(i)Q_i$, where $Q_i = \langle c_i^\dagger c_i \rangle$ and the weight $\text{w}(i) = 1$ for interior points and $\text{w}(i) = \frac{1}{4},\frac{2}{4},\frac{3}{4}$ for edge sites according to how the sites are ``shared" by internal and external regions [see Fig.~\ref{fig:disclination} (d)]. We determine the number of electrons filled in the system by requiring the system lie in the gap of the single-particle Hamiltonian before we switch on interaction (the implementation procedure is given in SI~\cite{suppl}). 

We first show the topological nature of the ground state. By threading an infinitesimal flux $\delta \phi$ in the plaquette near the disclination center, there should be $ 2\pi C\delta\phi$ charge pushed to the boundary according to Laughlin's argument based on Byers-Yang theorem~\cite{Laughlin1981Quantized}. In determining the Chern number, we choose the finite region with radius $R=3$ and measure the charge transfer across it. The Chern number can be computed by $C = \frac{n_W(\phi+\delta \phi) - n_W(\phi)}{\delta \phi}$. As depicted in Fig.~\ref{fig:disclination}(c), within the interaction range of $V<1.2$, the ground state remains within the IQH phase. The deviation of the quantization of the Chern number as $V$ increases, which we attribute to finite size effect. 


With the IQH phase in hand, we now examine the quantization of $\mathscr{S}_{\text{o}}$, which describes the charge response to disclinations.
The $\SO$ obtained from DMRG is present in Tab.~\ref{tab:SO and PO}, which quantizes to $1/2$ and matches prediction from Eq.~\ref{eq:S_C}. 
Region $W$ is choosed to enclose a single disclination with disclination angle $\Omega_W = \pi/2$.  
The excess charge in $W$ is defined as
\begin{equation}
    \delta Q = Q_W - \nu n_W ~~~\text{mod}~1 
    \label{eq:excess}
\end{equation}
where $n_W$ is the number of unit cells enclosed in region $W$ and $\nu$ is the background charge defined as the charge per unit cell in the bulk far from the defects and boundary. Upon our choice of the disclination center, $n_W$ should be an integer. Shift $\SO$ can be extract via 
\begin{equation}
\frac{\Omega_W}{2\pi}\SO = \delta Q.
\label{eq:discrete_shift_charge}
\end{equation}
We explore the evolution of $\delta Q$ with increasing interaction strength in Fig.~\ref{fig:disclination} (a). Starting from the regime without interaction, the DMRG result is consistent with the single particle calculation ~\cite{zhang2022fractional} $\delta Q=1/8$ (thereby $\SO = 1/2$ according to Eq.~\ref{eq:discrete_shift_charge}). The quantization of excess charge $\delta Q$ is converged for $R \geq 2$ as shown in Fig.~\ref{fig:disclination} (b). We choose $R=2$ in the interacting case Fig.~\ref{fig:disclination} (a), which is in the middle of the bulk as we turn on the interaction. $\delta Q$ converges to $1/8$ in the interacting regime up to $V=1.2$, yielding $\SO=1/2$. 
The competition between non-commutating kinetic term and interaction terms brings the quantum fluctuation, therefore when $V$ strength is comparable with $t$, the bond dimension $D$ should be increased to reach the quantization value. 
Our result extends previous works~\cite{zhang2022fractional,zhang2023quantized} on extracting crystalline symmetry protected topological invariants to correlated regime. 

Different from disclination, the dislocation can be defined on the cylinder, which reduces the edge effect and offers great convenience in DMRG simulations.
The result is $1/2$, also summarized in Tab.~\ref{tab:SO and PO}.
We choose the region enclosing the dislocation to be the cylinder away from the left \& right boundary and the defect as shown in Fig.~\ref{fig:dislocation}(d). The topological invariant $\PO$ can be obtained by measuring the excess charge trapped by the dislocation via 
\begin{equation}
\delta Q = \vec b \cdot \PO ,
\label{eq:excesscharge_Q_P}
\end{equation}
where $\vec b$ is the Burgers vector, which is $(0,1)$ in our study. The Chern number is determined by adiabatic flux insertion. Different from the disclination case, where the flux is inserted in the disclination center, for the dislocation we set a twist boundary condition $e^{i\theta}$ along the circumference of the cylinder. The charge will be pumped from the left boundary to the right boundary. As shown in Fig.~\ref{fig:dislocation}(c), the Chern number  $C=-1$ for $V<1.5$ recognizing that the system is in IQH phase. 

Polarization $\PO$ can be extracted by the excess charge bound to the dislocation, similar to the disclination case by using Eq.~\eqref{eq:excess}.
The dislocation introduces one extra irregular unit cell with area $\frac{1}{2}$, which should be counted in computing $\delta Q$(Eq.~\ref{eq:excess}). 
As depicted in Fig.~\ref{fig:dislocation}(a,b), $\delta Q$ converges to $\frac{1}{2}$ at bond dimension $D = 1200$ and for the region of width $4<R<6$. The polarization is $\mathscr{P}_{\OO,y}=\frac{1}{2}$ according to Eq.~\eqref{eq:excesscharge_Q_P}, consistent with the theoretical prediction. We note that the advantage of dislocation method is that this naturally gives an unambiguous polarization density even in a Chern insulator (see SI~\cite{suppl} for details). 

\begin{figure}
    \centering
    \includegraphics[width=1.0\linewidth]{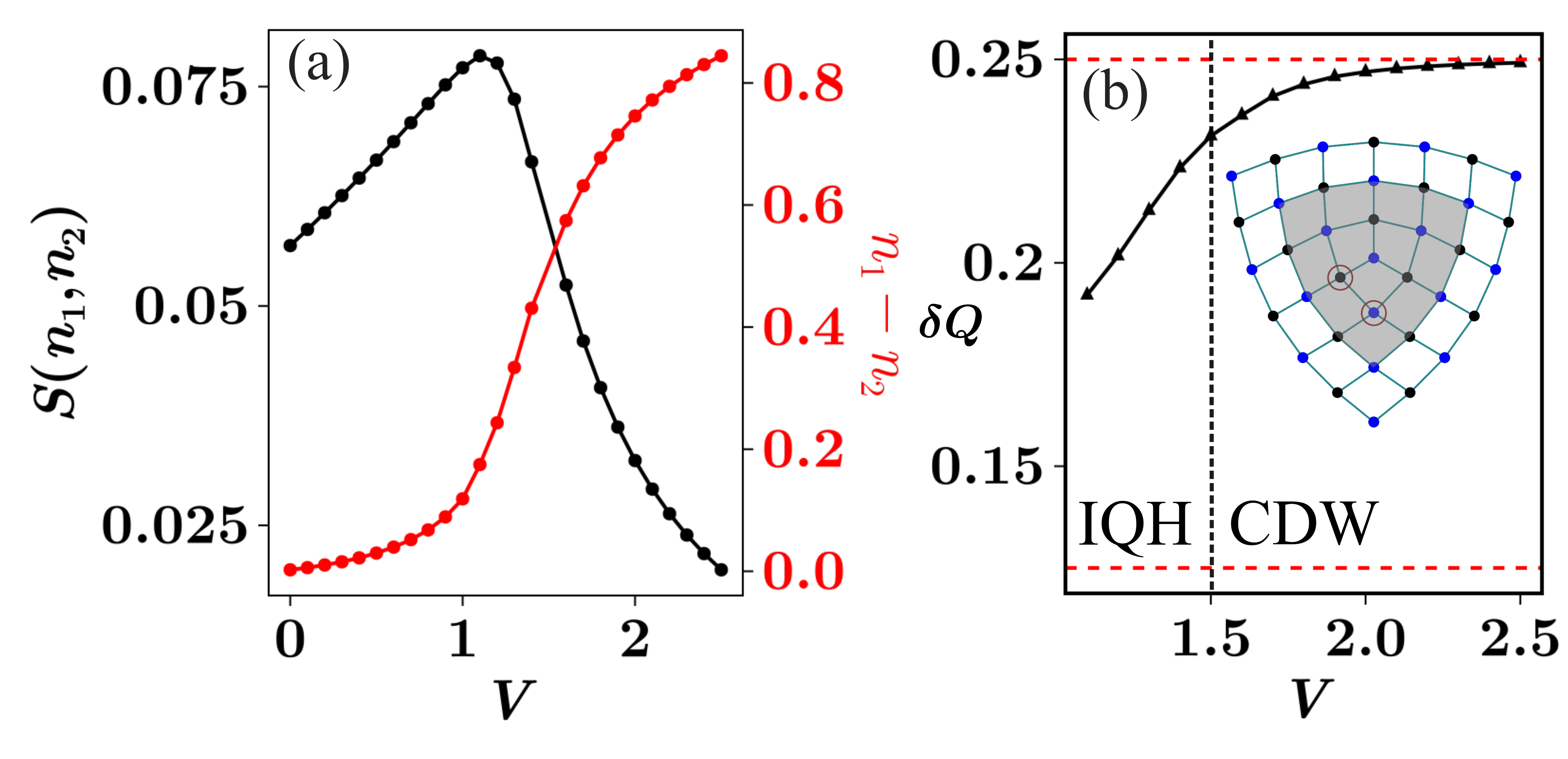}
    \caption{Transition from IQH to CDW as the interaction strength increases close to $\nu = \frac{1}{2}$. The IQH becomes unstable as the $V$ approaches the band gap. Panel (a) displays the evolution of charge occupation polarization $n_1 - n_2$ on 2 sites and the density-density correlation between them $\langle n_1 n_2\rangle - \langle n_1 \rangle \langle n_2 \rangle$ as we tune $V$. Panel (b) is the excess charge $\delta Q$ along this path.  The inset shows the CDW pattern, where the two sites encircled denote the position of $n_1$ and $n_2$ used in (a). The two red dashed lines highlight the quantization of $1/8$ and ${1/4}$.}
    \label{fig: cdw}
\end{figure}

\noindent{\textcolor{blue}{\it Strong interacting regime}---}
The strong repulsive interaction drives the system out of the IQH to a CDW phase as $V$ increases. We now tune the filling near $\frac{1}{2}$ to investigate the IQH-CDW transition, with $\phi = \pi$ to form a commensurate filling; while the Chern number of the system is still $-1$ in the weak interaction regime as before. The CDW pattern is shown in the inset of Fig.~\ref{fig: cdw} (b). We note that the original translation symmetry is spontaneously broken, but the $C_4$ rotation symmetry is still preserved in the CDW phase. Therefore, the system could still exhibit quantized discrete shift $\SO$ while the polarization $\PO$ needs to be defined with respect to the enlarged unit cell, 
i.e. $\PO$ can be extracted by constructing the dislocation with burgers vector $\vec b = (0,2)$. 
The result is shown in the SI~\cite{suppl}, where we have extracted vanishing excess charge bound to dislocations with burgers vectors $\vec b=(0,2)$. 
Fig.~\ref{fig: cdw}(b) displays the result of the excess charge near the disclination as the system enters the CDW phase from the IQH phase. We identify the transition by measuring the local order parameter of charge polarization $n_1 - n_2$ and local density-density correlation $S(n_1,n_2) = \langle n_1 n_2\rangle  - \langle n_1\rangle  \langle n_2 \rangle$ with site 1 and 2 encircled in red in Fig.~\ref{fig: cdw} (b). The result is shown in Fig.~\ref{fig: cdw} (a). 
The excess charge $\delta Q$ converges well to $0.25$ deep inside the CDW phase, which leads to the discrete shift $\SO = 1$.The disclination center can alternatively be located at the other sublattice, producing a vanishing discrete shift, which is discussed in the SI~\cite{suppl}. 
Therefore, the crystalline topological invariants have been fully determined in strongly interacting regime and align with the CDW pattern.
The integer discrete shift $\SO$ is consistent with the even Chern number $C$ according to Eq.~\eqref{eq:S_C}. 
This quantized discrete shift in the CDW side is a purely interaction-driven effect, which is beyond the single-particle band analysis.

Since the energy gap is close to the transition point, i.e. $V \sim 1.5$ in Fig.~\ref{fig: cdw} (b), the computation of $\delta Q$ requires very large size to reach quantization, which is beyond the scope of the present study. 

\noindent{\textcolor{blue}{\it Discussion}---}
Using DMRG, we observed that in an interacting system, both disclinations and dislocations can induce excess charge and give rise to the expected topological response. There exists finite-size effect in the former due to the open boundary, and the absence of it in the latter, as one can eliminate boundaries by imposing periodic boundary conditions along the $y$ direction.


A number of numerical and theoretical approaches have been proposed to detect symmetry-enriched structures in topologically ordered systems~\cite{pollmann2012detection,zaletel2016space,zaletel2017measuring,cincio2015classification,sunDynamical2018,huang2014detection,shiozaki2017many}, however, many of these methods rely on nonlocal operators that are difficult to implement within the matrix product state (MPS) framework.
For example, Refs.~\cite{zhang2022fractional,zhang2023quantized,ryu2010topological} utilized rotation operation to extract crystalline invariants, which is unfortunately challenging to be applied to cylinders with boundaries.
Our approach avoids non-local tensor network operation, which only involves measuring the local density. 
Although we focus on the fermionic interacting Hofstadter model in this paper, our approach can be analogously constructed in other systems such as spin model or bosonic systems. 
It has been recognized that the polarization $\PO$ defined as the charge response to the dislocation serves as a definition of absolute polarization, which is gauge invariant even if the Chern number $C$ is non-zero~\cite{gunawardana2025microscopic,zhang2025electric}. We provide a numerical validation of the definition in the interacting regime.

In conclusion, we numerically calculate the crystalline symmetry-protected topological invariant for interacting systems. The $\SO$ and $\PO$ are extracted by measuring the response of the IQH state to topological defects (disclination and dislocation). We also find the discrete shift is quantized in the CDW phase without an electron band.
Our study verifies the validity of the classification of the crystalline symmetry topological order in an interacting system. We also prove the feasibility of constructing symmetry defects in the framework of MPS. In addition to the theoretical interest, our result from the Hofstadter model can be directly verified in a cold atom or photonic system with symmetry defects~\cite{aidelsburger2013realization,miyake2013realizing,kennedy2015observation,hafezi2013imaging,ozawa2019topological}. 

Our work explores the interplay between $\text{U}(1)$ and crystalline symmetries in invertible phases. Looking forward, it is an on-going effort to employ the method developed here to study systems with intrinsic topological orders, e.g. fractional Chern insulators. It would also be interesting to investigate other symmetries or fractionalized phases, characterized by more complicated crystalline invariants beyond shift and polarization~\cite{kobayashi2024crystalline}. It is also promising to employ other numerical methods such as quantum Monte Carlo~\cite{heBonafide2016,heTopologicalI2016,sunDynamical2018} and $2$D tensor network, to explore the topological response of disclination and dislocation across the phase transition from a SET or SPT to a topologically trivial phase, where the Chern number computed via single-particle Green's function is found to be unable to capture the transition~\cite{heTopologicalII2016}.


{\it Acknowledgments}\,---\,  LZ and X-Y Song thank the helpful discussion with Maissam Barkeshli.
ML and ZYM acknowledge the support from the Research Grants Council (RGC) of Hong Kong (Project Nos. AoE/P-701/20, 17309822, HKU C7037-22GF, 17302223, 17301924), the ANR/RGC Joint Research Scheme sponsored by RGC of Hong Kong and French National Research Agency (Project No. A\_HKU703/22). We thank HPC2021 system under the Information Technology Services at the University of Hong Kong, as well as the Beijing Paratera Tech Corp., Ltd~\cite{paratera} and sugon cluster~\cite{sugon} for providing HPC resources that have contributed to the research results reported within this paper.

\bibliography{main}

\onecolumngrid 

\section{Supplemental Information}

\section{Field theory description of crystalline symmetry protected topological order}
The response Lagrangian in Eq.~(2) gives a compact way to describe charge pumping and defect-bound charge. 
Its physical meaning is similar to the Laughlin flux-insertion argument. 
There, inserting one unit of electromagnetic flux pumps a quantized charge. 
Here, the inserted flux can also be a ``crystalline flux``. 
The disclination acts as a flux of rotation symmetry, which has a simple physical meaning. 
When a particle moves once around the defect, it comes back to the same position, but the local lattice axes around it are rotated. 
This happens because a sector of the lattice has been removed or inserted at the defect core. 
Therefore, going around the defect is equivalent to applying a lattice rotation. 

We start from the mutual Chern-Simons term
\begin{equation}
    \frac{\SO}{2\pi} A\wedge dA_\omega .
\end{equation}
Here, $A$ is the external electromagnetic gauge field, and $A_\omega$ is the background gauge field for lattice rotations. 
a $2\pi$ flux of the spin connection $A_\omega$ represents a disclination defect.
Since $dA_\omega$ measures the disclination flux, this term states that a disclination carries electric charge. 
In particular, a $2\pi$ disclination binds charge $\SO$.

More explicitly, the electric current is obtained by varying the response action with respect to the electromagnetic field $A_\mu$ according to the linear response theory. 
Variation of the above term gives
\begin{equation}
    j^\mu
    =
    \frac{\delta S_{\rm response}}{\delta A_\mu}
    =
    \frac{\SO}{2\pi}
    \epsilon^{\mu\nu\rho}\partial_\nu A_{\omega,\rho}.
\end{equation}
Therefore, the total charge bound to a defect is
\begin{equation}
    Q
    =
    \int d^2x\, j^0
    =
    \frac{\SO}{2\pi}
    \int dA_\omega .
\end{equation}
For a $2\pi$ disclination, $\int dA_\omega=2\pi$, and hence $Q=\SO$. 
Thus, the coefficient $\SO$ can be measured from the excess charge near a disclination. 
This is the real-space response that we compute numerically in this work.

A similar idea applies to translation symmetry. 
A dislocation can be viewed as a flux of lattice translation. 

The response to this translation flux is controlled by the momentum polarization. 
Just as the disclination charge measures the response to a rotation flux, the charge or momentum response to a dislocation measures the response to a translation flux. 
In this sense, the momentum polarization can be understood as the topological coefficient that determines the response to a lattice-translation defect.

\section{Finite size and Finite bond dimension scaling}

\begin{figure}
    \centering
    \begin{overpic}[width=0.4\linewidth]{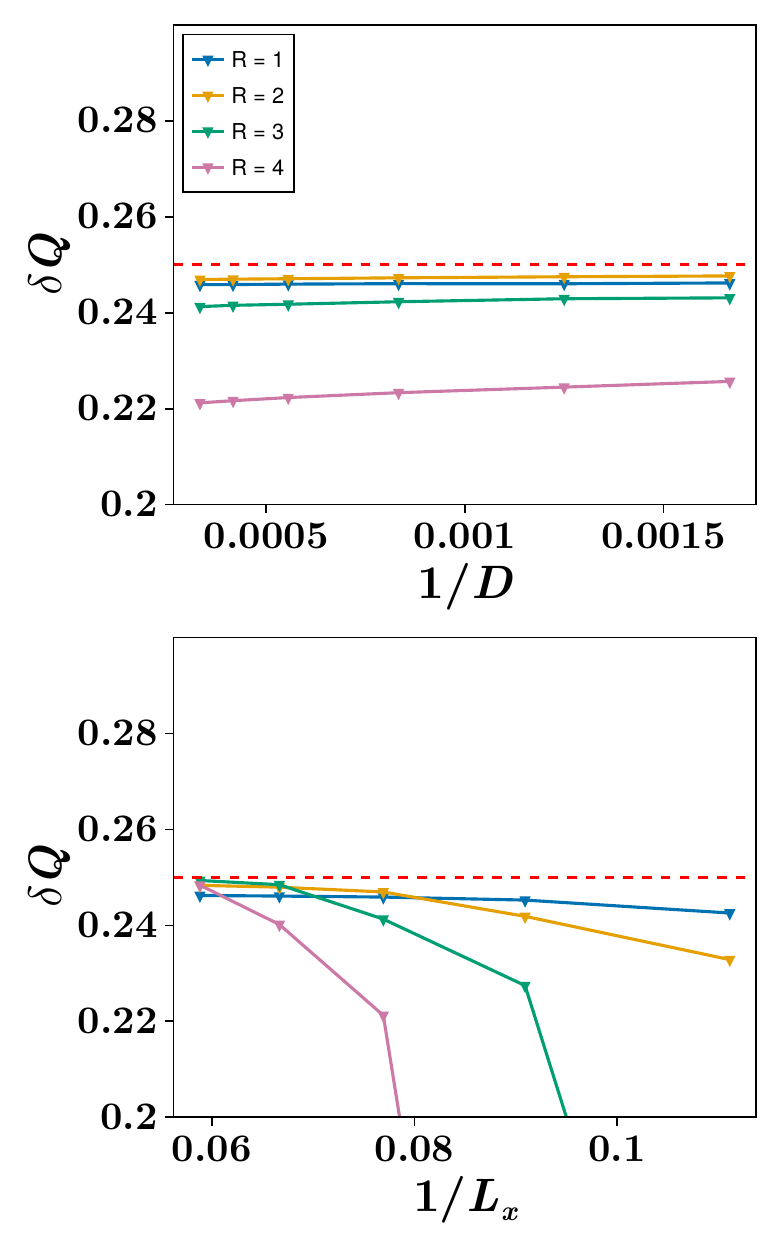}
        \put(55,94){(a)}
        \put(55,44){(b)}
    \end{overpic}
    \caption{\textbf{Finite bond dimension and finite size extrapolation for the discrete shift in the CDW phase($V = 2$)}. (a) Finite bond dimension extrapolation for different selected region. The calculation is performed on system with width $13$. The $R$ is the radius of the selected region. (b) Finite size extrapolation for bond dimension $3000$. $\delta Q$ is the excess charge near the symmetry defects}
    \label{fig:cdw_convergence}
\end{figure}

In this section we perform the finite-size and bond-dimension extrapolation of DMRG results for the discrete shift $\delta Q$, as shown in Fig~\ref{fig:cdw_convergence}. Fig~\ref{fig:cdw_convergence}(a) shows weak dependence on the inverse bond dimension 1/D for different region cut $L_xW$, indicating convergence in D. Fig~\ref{fig:cdw_convergence}(b)  shows clear finite-size scaling behavior as a function of $1/L_x$. The red dashed line indicates the predicted theoretical value $\delta Q = 0.25$. The main source of deviation from the prediction is identified as finite-size effects. The $L_xW = 2$ extrapolates to $0.25$ as $1/L_x = 0$, which we use to calculate the discrete shift over the whole parameter region.

\begin{figure}
    \centering
    \begin{overpic}[width=0.6\linewidth]{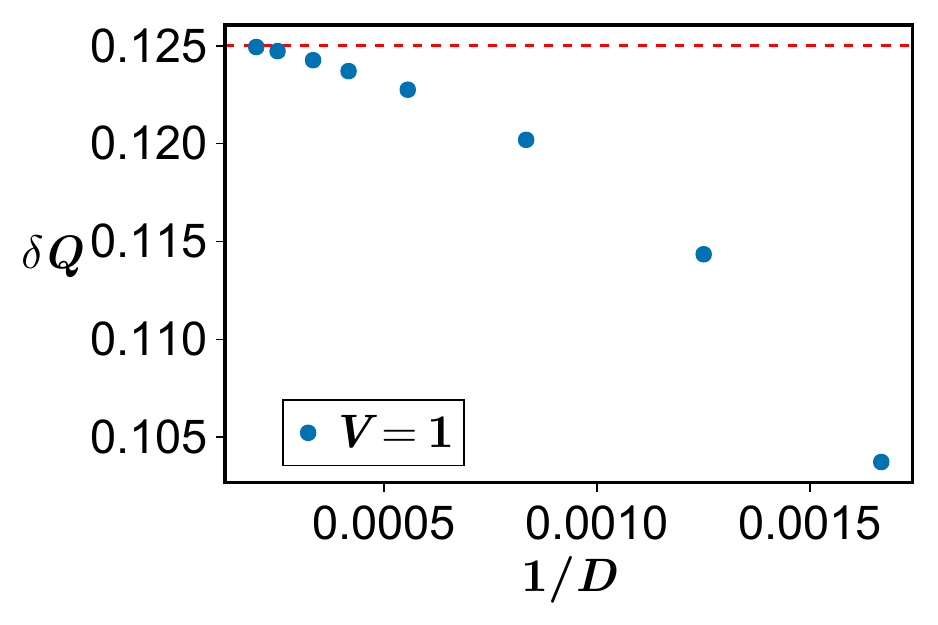}
    \end{overpic}
    \caption{\textbf{Finite bond dimension extrapolation for the discrete shift in the IQH phase($V = 1$)}.}
    \label{fig:iqh_convergence}
\end{figure}

Although the deviation from the quantized value of the discrete shift in the IQH phase is much smaller than that of the CDW phase, we also perform finite-bond-dimension scaling on this side. As shown in Fig~\ref{fig:iqh_convergence}, the quantization is recovered in better precision at $D = 5000$.

\section{Construction of the Hamiltonian in the presence with symmetry defects}

\begin{figure}
    \centering
    \includegraphics[width=0.6\linewidth]{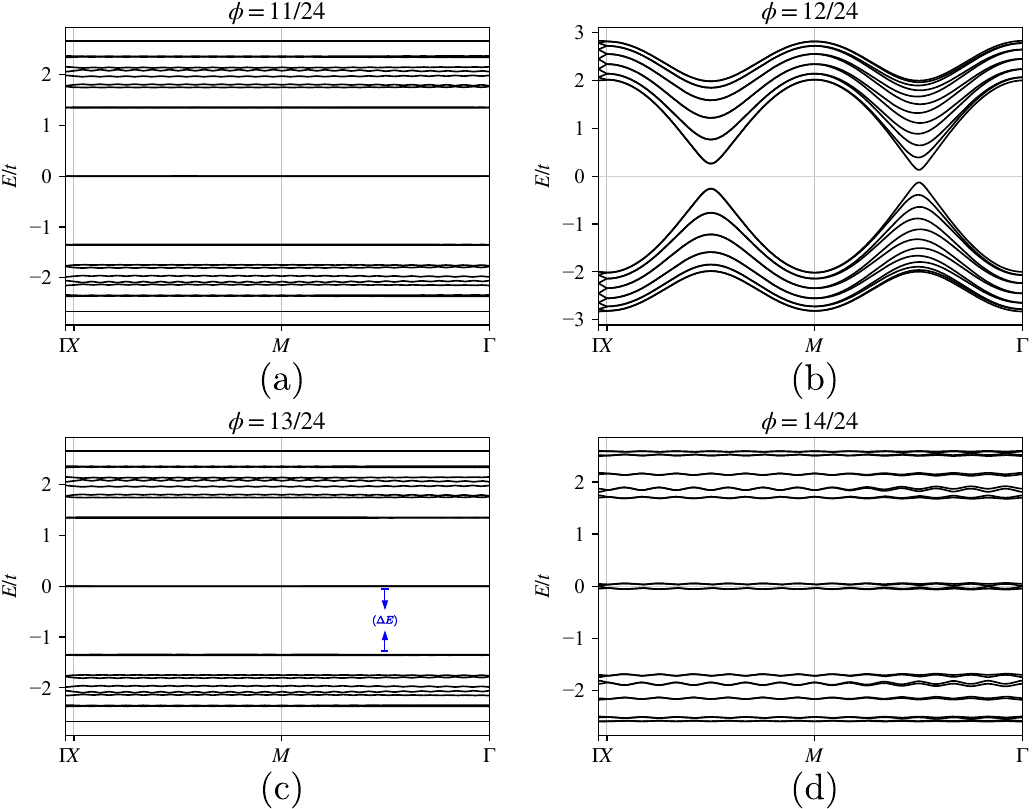}
    \caption{band structure of the Hofstadter model with flux $\phi = 11/24,12/24,13/24,14/24$. }
    \label{fig:band}
\end{figure}

In this section, we briefly explain how to construct the Hofstadter Hamiltonian with nearest neighbor interaction on square lattice with disclination and dislocation.
We start with the clean Hamiltonian, which is defined on a $L_x\times L_y$ lattice 
\begin{equation}
    H_{\text{clean}} = -t\sum_{\langle ij \rangle} e^{iA_{\text{clean},i,j}}c^\dagger_i c_j + V \sum_{\langle ij \rangle} n_i n_j +\mu \sum_i n_i,
    \label{eq:H}
\end{equation}
where $\langle ij\rangle$ sums over nearest neighbor terms. We choose Landau gauge where $\vec{A}_{\text{clean},x,y} = x\phi\hat {y}$. In our calculation, we set $L_x=L_y = L$ with open boundary conditions along both directions. 
The relation between the topological response and the filling of the Hofstadter Hamiltonian\eqref{eq:H} is
\begin{equation}
    \nu = \frac{C\phi}{2\pi}+\kappa,
\end{equation}
where $C$ is the Chern number of the system, $\kappa$ specifies the charge inside the unit cell and $\phi$ is the flux piercing each unit cell.
We plot the band structure of different flux per plaquette, shown in Fig.~\ref{fig:band}. The energy gap $\Delta E$ can be clearly seen.

\subsection{Lattice with disclination}
For $H_{\text{clean}}$, we can define magnetic rotation operator $\tilde C_4$ which commutes with $H_{\text{clean}}$(i.e. $[H_{\text{clean}},\tilde C_4] = 0$). $\tilde C_4 = e^{i\sum_j\lambda_j \hat n_j}C_4$ could be defined as ordinary $C_4$ operator followed by a gauge transformation $e^{i\sum_j\lambda_j \hat n_j}$, with $C_4$ defined as $C_4 a^\dagger_i C_4^\dagger =  a^\dagger_{C_4i}$. where $\hat n_i$ is the density operator and the $\lambda_i$ is the $\text{U}(1)$ phase defined on each site satisfying the condition:
\begin{equation}
A_{\text{clean},C_4 i, C_4 j}=A_{\text{clean},i j}+\lambda_j-\lambda_i. 
\label{eq:lambda}
\end{equation}
It can be viewed as the ordinary $C_4$ rotation followed by an onsite $\text{U}(1)$ transformation. This definition makes sure that the term $e^{iA_{\text{clean},i,j}}c^\dagger_i c_j$ transforms to the term $e^{iA_{\text{clean},C_4i,C_4j}}c^\dagger_{C_4i} c_{C_4j}$ upon the magnetic rotation. Thus, the Hamiltonian commutes with the magnetic rotation operator.

\begin{figure}[h]
    \centering
    \includegraphics[width=0.6\linewidth]{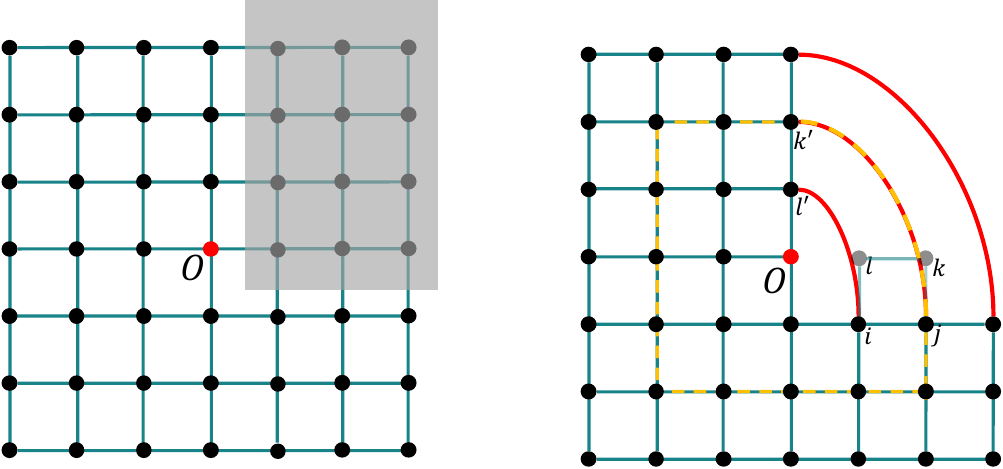}
    \caption{Cut-and-glue procedure for constructing the disclination with the center $O$, where $k'$ and $l'$ are the $C_4$ counterpart of $k$ and $l$. The sites inside the shaded area are removed and the red bonds are connected.}
    \label{fig:fig_cut_glue_SI}
\end{figure}

To construct the disclination, we first determine the $C_4$ rotation center $O$ of the square lattice, marked by the red dot. In our study, we chose the rotation center located at the vertex, which restricts $L$ to be odd to make the lattice $C_4$ symmetric. Then, we remove the sites in the shaded area and connect the dangling bonds. The gauge $A_{il'}$ connecting the sites on the edge of the shaded area is determined such that the flux $\phi_{ijkl} = A_{ij}+A_{jk'}+A_{k'l'}+A_{l'i}$ of new plaquette is the same as the flux $\phi_{ijkl}$ of $H_{\text{clean}}$. By equating $\phi_{ijkl} = \phi_{ijk'l'} $ we have:
\begin{equation}
    A_{jk}+A_{kl}+A_{li} =A_{jk'}+A_{k'l'}+A_{l'i}
    \label{eq:equality of phase}
\end{equation}
According to the definition of $\lambda_i$: $A_{k'l'}=A_{kl}+\lambda_k - \lambda_l$, the equation \eqref{eq:equality of phase} can be reduced:
\begin{equation}
    A_{jk}+A_{li} =A_{jk'}+A_{l'i}+\lambda_k - \lambda_l
\end{equation}
Thus the simplest choice of $A_{jk'}$ and $A_{il'}$ is
\begin{equation}
\begin{aligned} 
A_{jk'} = A_{\text{clean},jk}-\lambda_k  \\
A_{il'} = A_{\text{clean},il}-\lambda_l
\end{aligned}
\end{equation}
And we finish the cut-and-glue procedure for the hopping terms. For the interaction terms, $n_i n_{l}$ simply becomes the $n_i n_{l'}$ under the cut-and-glue procedure.

 This approach of determining the gauge field has the global $\text{U}(1)$ phase ambiguity. In the following, we show that we can utilize this ambiguity to perform the flux insertion, which is used to determine the Chern number of the system in our study. Suppose we add a constant $\phi_0$ to all the $\lambda_i$ defined site by site. The equation \eqref{eq:lambda} is still preserved since the over all constant $\delta \phi$ of $\lambda_i$ is canceled by the subtraction term $\lambda_i -\lambda_j$. However, there will be additional $\delta \phi$ flux accumulated around the loop(orange dashed line in the Fig.~\ref{fig:fig_cut_glue_SI}) after the cut-and-glue procedure. This is equivalent to inserting a flux into the center of the disclination. If the Chern number of the system is non-zero, there should be $Cd\phi$ charge pumped from the center to the boundary.

\subsection{Lattice with dislocation}

\begin{figure}
    \centering
    \includegraphics[width=1.0\linewidth]{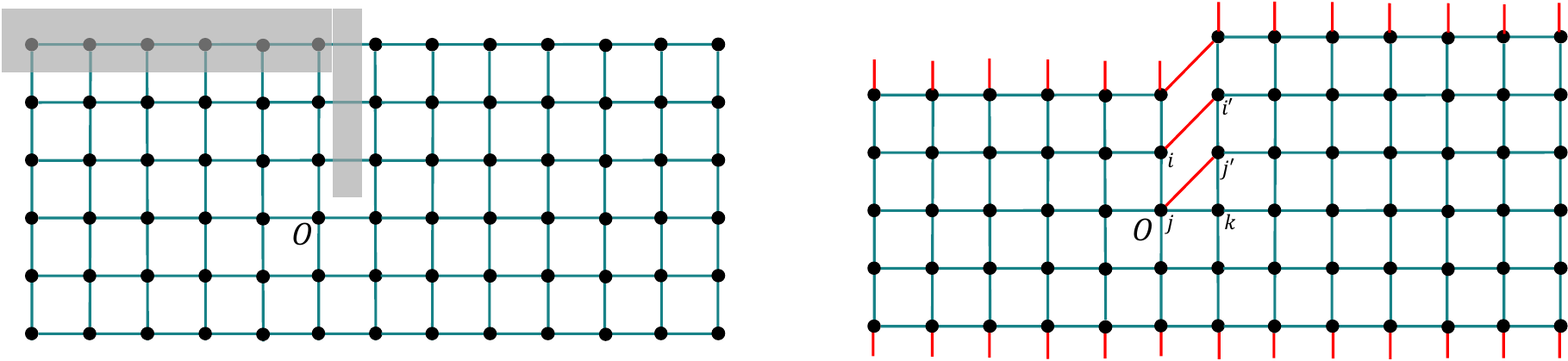}
    \caption{cut-and-glue procedure for constructing the dislocation. The procedure is same as the disclination with sites in the shaded area removed and red bonds connected.}
    \label{fig:fig_cut_glue_SI}
\end{figure}

To construct the dislocation, we start from the Hamiltonian defined on the clean lattice and perform a similar cut-and-glue procedure by removing the sites and bonds covered by the shaded area and connecting the red bonds.
On the clean lattice, the magnetic translation symmetry is defined as $\tilde T_x = T_x e^{i\sum_j\hat n_j \lambda_j^x}$ with the $\lambda^x_j$ defined as
\begin{equation}
    A_{T_x i, T_x j}=A_{i j}+\lambda_j^x-\lambda_i^x.
\end{equation}
The definition of $\tilde T_y$ is the same as $\tilde T_x$. Now we can make use of the $\lambda_{x(y)}$ to determine the gauge connecting the bonds. To make sure the phase of the new plaquette $\phi_{ijj'i'}$ from cut-and-glue is the same as others, the gauge field $A_{jj'}$ and $A_{ii'}$ is
\begin{equation}
    \begin{aligned}
        A_{ii'} =  A_{\text{clean},ij'} - \lambda_{k }^y\\
        A_{jj'} =  A_{\text{clean},jk}  - \lambda_{j'}^y\\
    \end{aligned}
\end{equation}
from the same proof in the case of disclination. In this case, we fix the overall phase ambiguity by setting $\lambda^y_i$ at the point $O$ to be zero. There is no net flux going around the dislocation center. 
To eliminate the boundary effect, we also connect the sites along the $y$ direction to impose the periodic boundary condition. Under the Landau gauge choice with only $A_{i,i+\hat x}$ non-zero, the gauge connecting the two edges is simply $0$. However, it is non-zero when inserting the flux through the cylinder.

\section{Details of DMRG calculation}
 In our simulation of disclination, we consider a square lattice with $13\times 13 $ unit cells, which is cut-and-glued to introduce a disclination defect with disclination angle $\pi/2$. The total number of sites after the cut-and-glue is $127$. 

\begin{figure}
    \centering
    \includegraphics[width=1.0\linewidth]{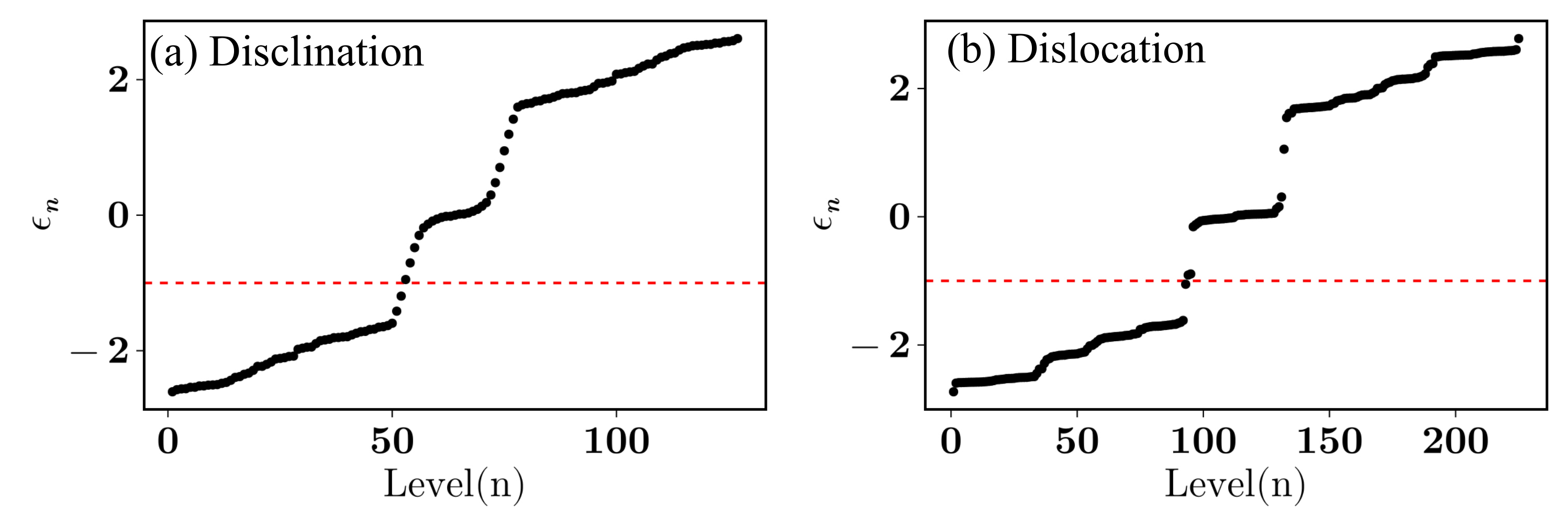}
    \caption{Single particle level of the cluster with disclination(a) and dislocation(b). The red dashed line shows the chemical potential $\mu = -1$ we choose.}
    \label{fig:fig_single_level}
\end{figure}

For this irregular geometry, the MPS chain is arranged as shown in Fig. 1(a) of the main text. To determine the number of electrons, we check the single-particle energy level. As depicted in Fig~\ref{fig:fig_single_level}. The chemical potential is chosen such that the bulk region of the cluster is gapped. While the gapless states near $\mu$ belong to edge states that will not affect the excess charge. 

Charge $\text{U}(1)$ symmetry is implemented in our DMRG calculation based on the ITensor package~\cite{ITensor,ITensor-r0.3}. We keep up to $m=3000$ states in the DMRG simulation for disclination to ensure that the maximum truncation error is around $10^{-4}$. The MPS is sensitive to the width. Our calculation shows it is already sufficient to obtain the discrete shift to a great precision. For dislocation, we consider a square lattice with $8\times 30 $ unit cells and perform cut-and-glue procedure. The total number becomes $225$. Periodic boundary conditions along the y-direction can be applied in the case of dislocation, where the boundary states are eliminated, which reduces the size effect significantly. The remaining sites after cut and glue are $150$. We keep up to $m=1800$ states in the DMRG simulation for disclination to ensure that the maximum truncation error is around $10^{-6}$.

\section{Disclination and dislocation with random interaction}
\begin{figure}
    \centering
    \includegraphics[width=1.0\linewidth]{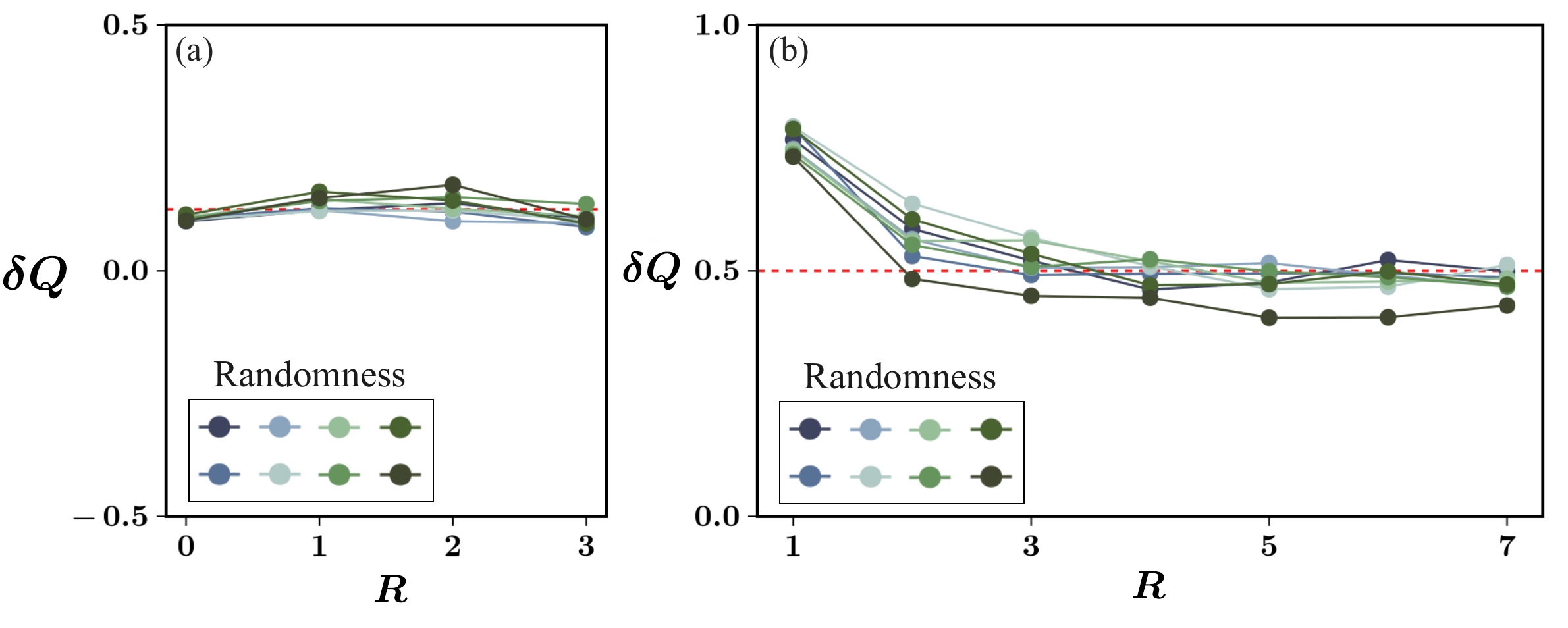}
        \caption{The excess charge around the disclination(a) and dislocation(b) by adding the random nearest interaction. The interaction strength is the uniform distribution from $0$ to $0.6$ in both cases. Different lines represents the random interaction with different seeds.}
    \label{fig:randomness}
\end{figure}

To further explore the role of interaction, we calculate the excess charge when the strength of the interaction terms is random. 
In this case, the Hamiltonian breaks the crystalline symmetry, and the bulk wave function is not homogeneous. 
The excess charge $\delta Q$ will fluctuate near the quantization value and not fully break down, showing the robustness of our result~\ref{fig:randomness}.
This phenomenon can be understood pictorially.
When the interaction term breaks the symmetry, the electrons push each other in a non-uniform way, resulting in a non-uniform charge density distribution.
Thus, the total charge inside the region $W$ fluctuates, and the excess charge inside the region $W$ also differs from the average value.

\section{The polarization density in a Chern insulator is well-defined}
\label{sm:welldefined}
There has been ambiguity in the polarization density in phases with nonzero Chern number. One simple picture is that given a Chern insulator on a cylinder, an adiabatic change of the gauge connection uniformly as $A_i\rightarrow A_i+2\pi/L$, along the circumference of the cylinder will pump $C$(Chern number) charges from one end to the other end. The system after the adiabatic evolution, is equivalent to the original one by a large gauge transform. The charge pumping, however, will change the polarization density by $C/L$ where $L$ is the circumference.  We emphasize that in our scheme of extracting polarization density from dislocation, this ambiguity is naturally circumvented. To demonstrate this, below is the excess charge in the non-interacting Hofstadter model upon a large gauge transform in the transverse direction, which remains nicely quantized to the expected polarization value. The physical picture is that the charge pump changes the total polarization, nevertheless the polarization in the bulk remains invariant which a dislocation in the bulk is capable of probing.

\begin{figure}
    \centering
    \includegraphics[width=1.0\linewidth]{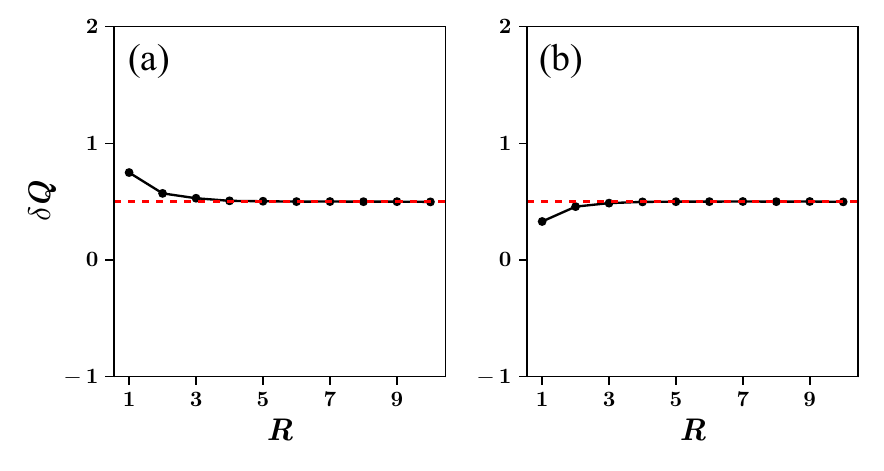}
    \caption{The excess charge obtained from single particle calculation. The figure shows the excess charge versus the region width in the original Hamiltonian(a) and the Hamiltonian with large gauge transformation(b) that pumps the unit charge from left to right edges. The excess charge both converges to the same value.}
    \label{fig:polarization}
\end{figure}

\newpage

\section{The electric polarization and discrete shift in the CDW phase}

\begin{figure}
    \centering
    \includegraphics[width=1.0\linewidth]{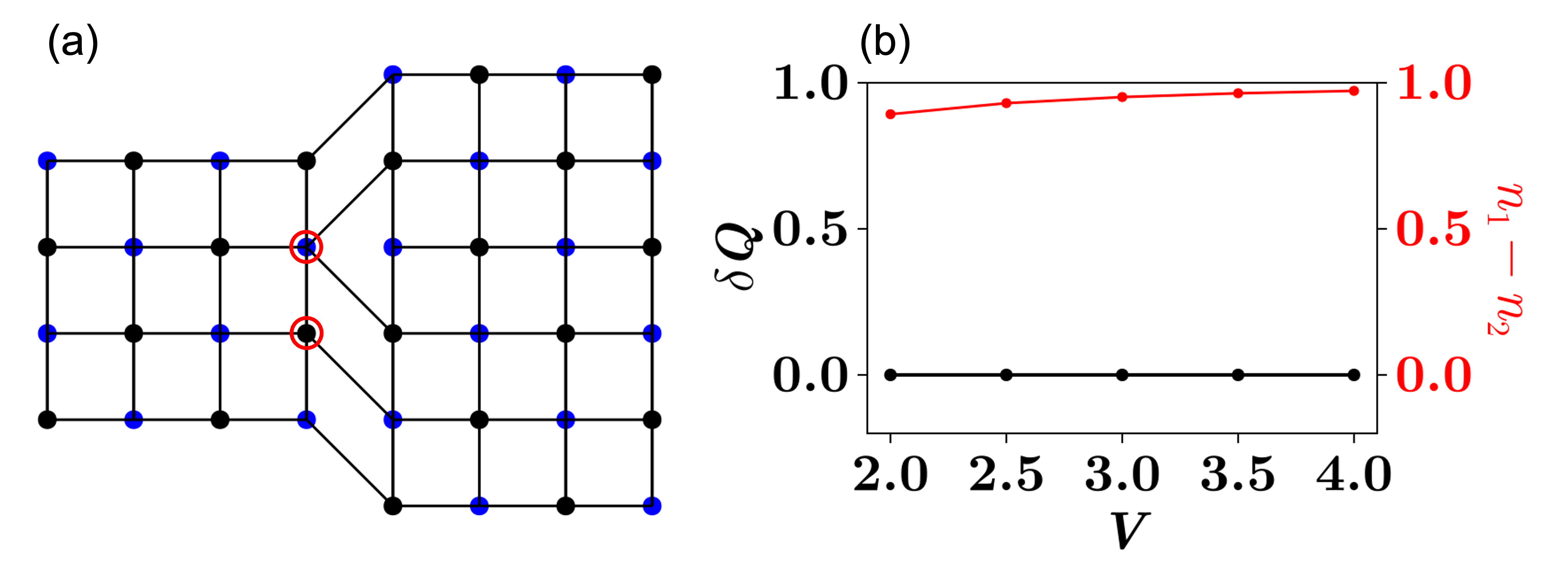}
    \caption{CDW phase of Hofstadter square lattice model cylinder with dislocation $b = 2$ defect at large $V$ limit. The system size is $L_x\times L_y = 30\times 12$. The flux per plaquette is $1/2$ and we fully fill the lowest band to construct a commensurate CDW, with filling $1/2$. Panel (a)  displays the CDW pattern, where the two sites encircled in red denote the position of $n_1$ and $n_2$ used in measuring charge occupation polarization $n_1 - n_2$ on (b), Here we use $L_x = 8, L_y = 6$ for demonstration. (b) is the excess charge $\delta Q$ and charge polarization obtained via DMRG deep inside the CDW phase.}
    \label{fig:polarization_cdw}
\end{figure}

The polarization in the translation symmetry breaking phase can be captured by constructing the "larger" dislocation~\ref{fig:polarization}(b), with burgers vector $(0,2)$ for convenience. 
The Hamiltonian is similarly constructed via cut-and-glue procedure while the interacting strength should be sufficient large compared with the kinetic term to render the system in the CDW phase. The charge distribution is shown in Fig.~\ref{fig:polarization_cdw}(a).
The excess charge remains zero for $2.0<V<4.0$ from Fig.~\ref{fig:polarization_cdw}(b), which corresponds to the vanishing polarization $\mathscr{P}_{o,y}$, according to $\delta Q = \vec b \cdot \PO$. 

The discrete shift in the CDW phase has been shown in the main text. Because of the translation symmetry breaking, there are two sublattices in the CDW phase. It is natural that the disclination center is located at the other sublattice. From the previous work, the discrete shift $\SO$ with different disclination center $o$ can be related with the following simple formula~\cite{zhang2023quantized}:
\begin{equation}
   \mathscr{S}_{\beta}=\mathscr{S}_{\alpha}+2 \mathscr{P}_{\alpha,y} - \kappa
\end{equation}
where $\alpha$ and $\beta$ denote different disclination center. In the CDW phase, $\kappa$ is the charge in each unit cell, which is equal to $1$. It can be seen directly that the discrete shift $\SO$ at the other sublattice is $0$ with vanishing polarization $\mathscr{P}_{\alpha,y}$.
Now consider the fully polarized CDW phase with the electrons only occupying one sublattce and the disclination center is not occupied by any electrons in the CDW limit.
\begin{figure}[h]
    \centering
    \includegraphics[width=0.5\linewidth]{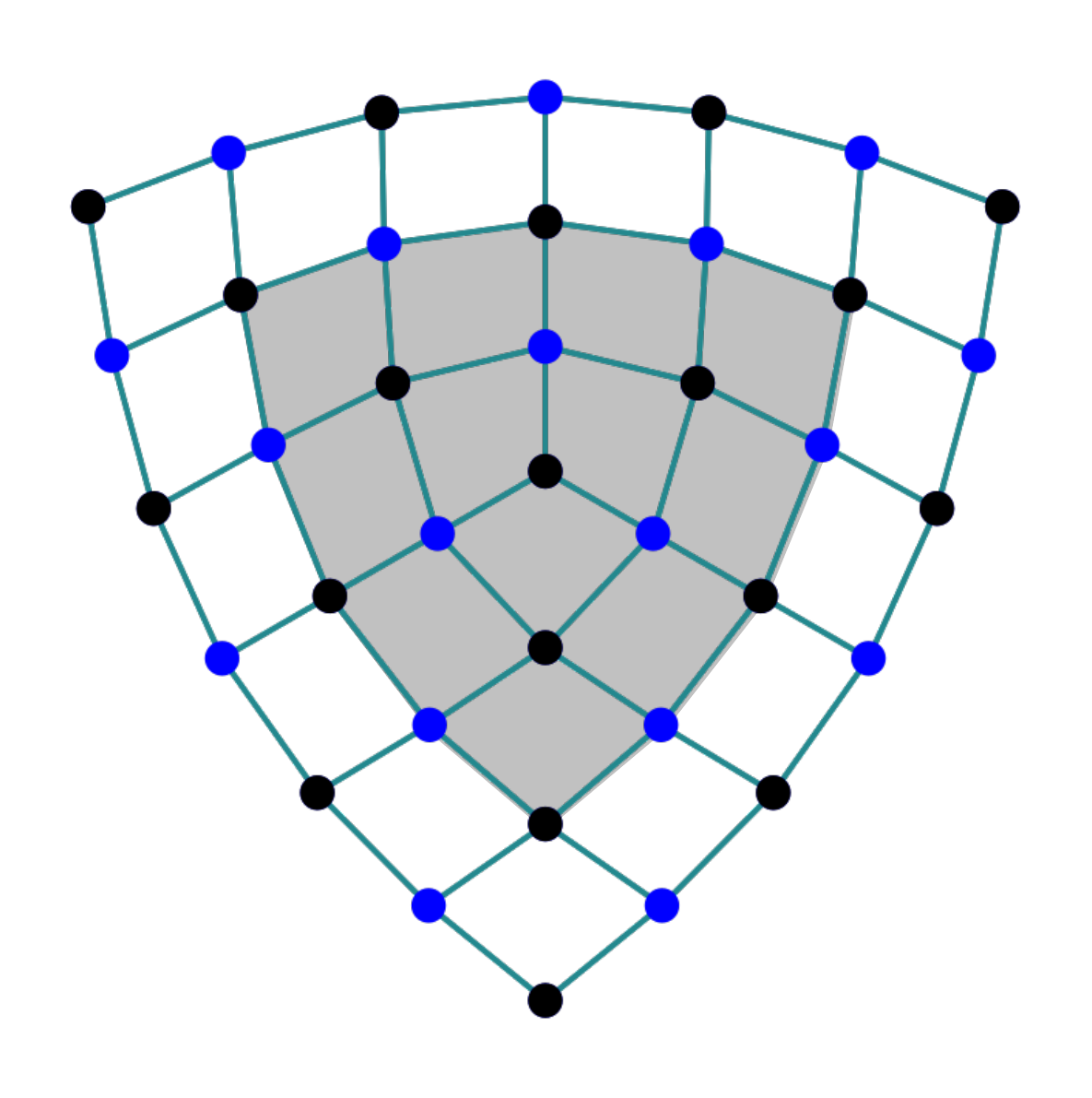}
    \caption{The fully polarized CDW phase with disclination. The disclination center is unoccupied}
    \label{fig:placeholder}
\end{figure}

There are $12$ unit cells in the shaded area and $\nu = 1/2$. There are $6$ electrons in the region, which leads to the $0$ excess charge. Thus in the CDW phase with different disclination center, $\mathscr{S}_{\beta} = 0$.

\end{document}